# From Qubits to Qumodes: Information Capacity of Anyonic Excitations

Satish Prajapati[*, †]
Department of Ceramic Technology, Government College of Engineering and Ceramic Technology,
Kolkata, West Bengal 700010, INDIA

**ABSTRACT**. The interplay between quantum statistics and information encoding is a cornerstone of quantum physics. Here, the maximum information capacity of a quantum state governed by Haldane's exclusion statistics is derived. The capacity, defined by the maximum von Neumann entropy of its occupancy distribution, follows $S_{\max}(g) = \log_2(\lfloor 1/g \rfloor + 1)$. This result continuously interpolates between the fermionic limit of a single qubit ($g = 1$) and the bosonic limit of a continuous-variable qumode ($g \to 0$) For the $\nu = 1/3$ fractional quantum Hall state ($g = 1/3$), we predict a 2-bit capacity, observable as four distinct quantized conductance plateaus in quantum dot spectroscopy, providing a direct signature of anyonic statistics.

## I. INTRODUCTION.

Quantum statistics fundamentally govern the behavior of identical particles, dividing them into two classes in three dimensions: fermions and bosons. Fermions obey the Pauli exclusion principle, a behavior captured by Haldane's statistical parameter $g = 1$, allowing at most one particle per quantum state. Bosons ($g = 0$) permit unlimited occupancy. This distinction has profound implications for quantum information encoding, where fermionic states naturally serve as qubits (1-bit capacity) and bosonic states as qumodes, systems with unbounded information capacity used in continuous-variable quantum information [1].

In two-dimensional systems, particularly in the fractional quantum Hall effect (FQHE), anyons exhibit intermediate statistics characterized by a continuous parameter g [2,3]. While the thermodynamic properties of anyonic gases have been extensively studied [4, 5], a fundamental question remains: how does the continuous nature of anyonic statistics directly govern the information-carrying capacity of a single quantum state? Answering this question would bridge a conceptual gap between topological matter and quantum information science, complementing prior work on topological quantum computation [6].

In this work, we derive the maximum information capacity: $S_{\max}(g) = \log_2(\lfloor 1/g \rfloor + 1)$.
We show that $S_{\max}(g)$ evolves in a quantized manner from 1 bit to infinity as g varies from 1 to 0, effectively connecting the qubit and qumode paradigms. We further propose experimental signatures through quantized conductance measurements in quantum dot spectroscopy of FQHE systems.

## II. THEORY

### A. Haldane Exclusion Statistics

Haldane's exclusion statistics generalizes the Pauli principle through a statistical parameter g [2]. For a single quantum state, the maximum number of particles $m$ that can occupy it is given by: $m = \lfloor 1/g \rfloor$. where $g = 1$ for fermions and $g = 0$ for bosons. This definition leads to: $g = 1/2 (semions) : m = 2$, $g = 1/3 : m = 3$, $g = 1/4 : m = 4$.

### B. Grand Canonical Partition Function

For a system of non-interacting particles obeying exclusion statistics, the grand canonical partition function factorizes over independent single-particle states. While the underlying electron system in the FQHE involves strong Coulomb interactions, the low-energy excitations are emergent quasiparticles. Haldane's exclusion statistics provides an **effective,** non-interacting description for these quasiparticles, precisely capturing their statistical properties in the low-energy limit where interaction effects are renormalized into the statistical parameter $g$ [4, 7]. This justifies the use of the single-state partition function formalism for calculating the occupancy probabilities of these topological excitations. For a single state at energy $\varepsilon_i$, the partition function sums over all allowed occupancies:

$$Z_i = \sum_{n_i=0}^{m} e^{-\beta n_i(\varepsilon_i - \mu)} = \frac{1 - e^{-\beta(m+1)(\varepsilon_i - \mu)}}{1 - e^{-\beta(\varepsilon_i - \mu)}}$$

where μ is the global chemical potential and $\beta = 1/(k_B T)$. This general form reduces to known cases:
Fermions $(g = 1, m =) : Z_F = 1 + e^{-\beta(\varepsilon_i - \mu)}$
Bosons $(g = 0, m \to \infty): Z_B = \frac{1}{1 - e^{-\beta(\varepsilon_i - \mu)}}$

Semions ($g = 1/2, m = 2$): $Z_{1/2} = 1 + e^{-\beta(\varepsilon_i - \mu)} + e^{-2\beta(\varepsilon_i - \mu)}$.

Information Capacity via Von Neumann Entropy

The probability of occupancy $n$ is given by the Gibbs distribution:

$$P(n) = \frac{e^{-\beta n(\varepsilon_i - \mu)}}{Z_i}$$

The von Neumann entropy for this distribution is:

$$S(g) = -\sum_{n=0}^{m} P(n) \log_2 P(n)$$

The maximum entropy $S_{\max}(g)$ is achieved when all occupational states are equally probable, which occurs when the energy level aligns with the chemical potential ($\varepsilon_i = \mu$). This condition is precisely what is scanned through when sweeping a gate voltage $V_g$ in a quantum dot transport experiment, making the maximum entropy regime directly accessible [4, 8]:

$$P(n) = \frac{1}{m+1} \quad \text{for all } n = 0,1,\dots,m.$$

Substituting into the entropy formula gives the maximum information capacity:

$S_{\max}(g) = -\sum_{n=0}^{m} \frac{1}{m+1} \log_2\left(\frac{1}{m+1}\right) = \log_2(m+1)$

Expressing this in terms of the statistics parameter $g$:

$$S_{\max}(g) = \log_2(\lfloor 1/g \rfloor + 1)$$

This maximum entropy, $S_{\max}(g)$, represents the classical information capacity—the number of bits that can be reliably stored and read out via a projective charge measurement of the quantum state's occupancy.

## III. RESULTS

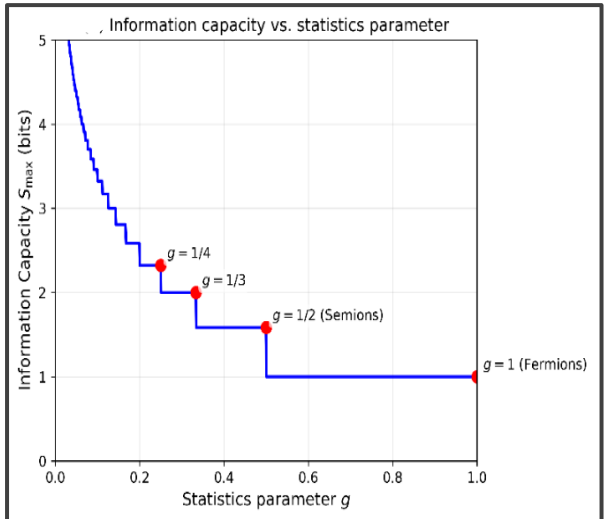

FIG. 1. Information capacity and occupancy distributions. Maximum information capacity $S_{\max}$ of a single quantum state as a function of the exclusion statistics parameter $g$. The quantized staircase function interpolates between a fermionic qubit (1 bit at $g = 1$) and a bosonic qumode (infinite capacity as $g \to 0$). As shown in Fig. 1, $S_{\max}(g)$ versus $g$ forms a quantized staircase, highlighting the transition from qubit to qumode: $g = 1: S_{\max} = \log_2(2) = 1 bit (fermionic\ qubit), g = 1/2: S_{\max} = \log_2(3) \approx 1.585 bits$, $g = 1/3: S_{\max} = \log_2(4) = 2 bits, g = 1/4: S_{\max} = \log_2(5) \approx 2.322 bits$, $g \to 0: S_{\max} \to \infty (bosonic\ qumode)$. This function defines a continuous transition from discrete digital information(qubits) to continuous analog information (qumodes) governed solely by quantum statistics.

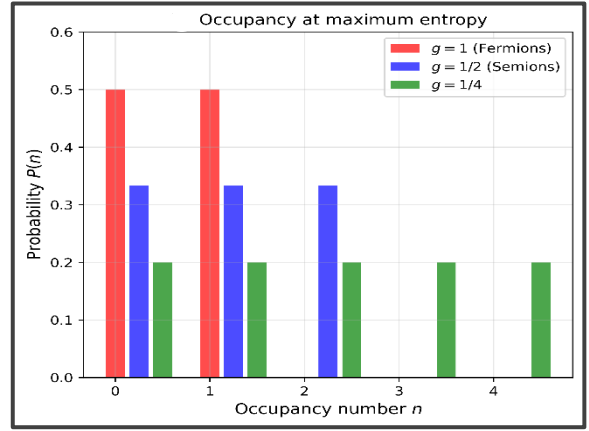

FIG 2. The probability distribution ($P(n)$ of occupation number n at maximum entropy for three specific cases: fermions ($g = 1, m = 1$), semions ($g = 1/2, m = 2$), and a case with ($g = 1/4, m = 4$). The capacity is the $S = -\sum_n P(n) \log_2 P(n)$ of these distributions.

### A. Experimental Implications Quantum Dot Spectroscopy

For the $\nu = 1/3$ fractional quantum Hall state ($g = 1/3$), we predict a quantum dot can trap $n = 0, 1, 2, 3$ anyons [8]. As depicted in Fig. 2, the tunneling conductance through such a dot should exhibit four distinct quantized plateaus, corresponding to $n = 0, 1, 2, 3$ anyon occupancies. For quasiparticles of charge $q = e/3$, these plateaus are expected at conductances quantized at values proportional to $n \cdot q^2/h = n \cdot (e^2/9h)$, for $n = 0, 1, 2, 3$ [8], directly measuring the 2-bit information capacity. While the signal for $n = 1$ ($\sim e^2/9h$) is small, modern ultra-low-noise measurement techniques at millikelvin temperatures have successfully resolved such quantized states in FQHE dots [9]. This measurement, feasible at millikelvin temperatures and magnetic

*Satish Prajapati: ORCID iD: **0009-0006-3801-1137**
†Satish Prajapati: iamsatish.gcect.ac@gmail.com

fields of ~5–10 T in GaAs-based quantum dots [8, 9], requires the charging energy of the dot $E_C$ to satisfy $E_C \gg k_B T$ to overcome thermal broadening. High-resolution gate control is also essential to resolve the discrete anyon occupancies against disorder-induced energy scales. This signature should be accompanied by characteristic shot noise modulation at plateau transitions, reflecting the $e/3$ quasiparticle charge [10,11]. For non-abelian anyons (e.g., at $\nu = 5/2$), the capacity may differ due to braiding statistics, requiring further theoretical exploration [6]. These predictions are testable using momentum-resolved tunneling spectroscopy [14,15], which can probe quasiparticle occupancies in quantum dots by detecting tunneling currents, or Fabry-Pérot interferometry [12,13] for complementary edge-state measurements.

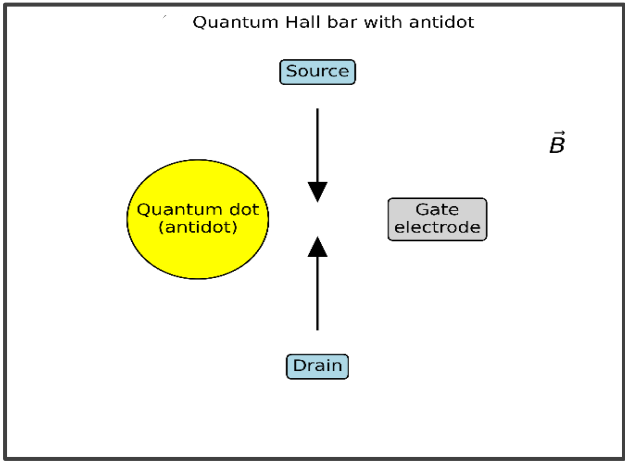

FIG. 3. Proposed experimental signature in quantum dot spectroscopy. Schematic of a quantum Hall bar device. A gate-defined antidot (or quantum dot) traps anyonic quasiparticles in a fractional quantum Hall state (e.g., $\nu = 1/3$).

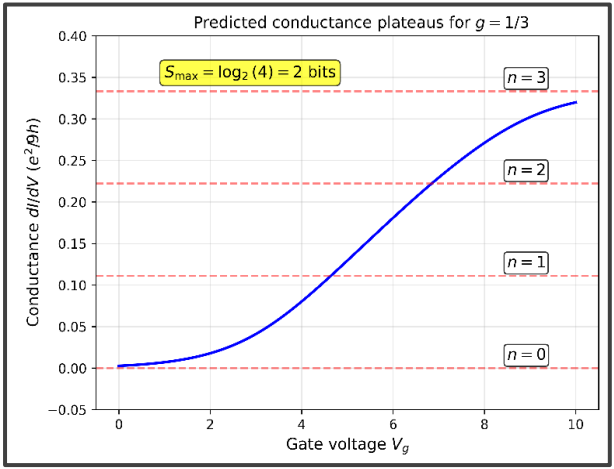

FIG 4. Predicted low-temperature differential conductance ($dI/dV$ as a function of gate voltage $V_g$. For statistics parameter $g = 1/3$ (max occupancy $m = 3$)), four distinct plateaus are predicted, corresponding to quantized tunneling through the dot occupied by ($n = 0, 1, 2, 3$) anyons. The plateau conductances are proportional to $n \cdot (e^*)^2$ (where $e^* = e/3$), providing a direct measurement of the $(\log_2(4) = 2$ bit information capacity.

## IV. DISCUSSION

While the mathematical derivation of $S_{\max}(g) = \log_2(m+1)$ is straight forward, its physical implication is profound: it defines a universal, statistics-dependent information capacity for a quantum state. The key insight is not the calculus of entropy itself, but the synthesis of Haldane's exclusion principle with information theory to create a quantitative metric that connects abstract quantum statistics to concrete experimental observables. This formalism is general. For example, beyond the $\nu = 1/3$ state, at $\nu = 1/5$ (where $g = 1/5$ and $m = 5$), a quantum dot should exhibit six distinct conductance plateaus, corresponding to a maximum classical information capacity of $\log_2(6) \approx 2.58$ bits. For multi-state systems, the total capacity would scale with the number of independent states, potentially enabling high-dimensional encoding in anyonic quantum memories, as explored in topological quantum computation [6]. The capacity $S_{\max}(g)$ generalizes the concept of a qubit to a "statistics-tunable" information carrier. For anyonic systems, this reveals that a single quantum state possesses an intrinsic higher-dimensional information capacity (e.g., a 4-level system for $g = 1/3$), fundamentally extending the binary encoding offered by fermionic qubits. This suggests a novel pathway toward realizing native qudits for quantum simulation, where a 4-level system ($g = 1/3$) could enable compact encoding for quantum error correction or simulation of topological quantum field theories [6]. The predicted conductance plateaus provide a directly testable signature in existing FQHE platforms [8, 9]. Deviations from the ideal staircase behavior could reveal effects of electron-electron interactions beyond the exclusion statistics paradigm or provide evidence for non-abelian statistics in other filling fractions.

## V. CONCLUSIONS

We have derived the maximum information capacity of a quantum state under exclusion statistics, showing it follows a quantized staircase function $S_{\max}(g) = \log_2(\lfloor 1/g \rfloor + 1)$. This work unifies the qubit and qumode paradigms through the mechanism of statistical transmutation and proposes concrete experimental verification via quantized conductance measurements in anyonic quantum dot spectroscopy. More broadly, our formalism provides a quantitative metric for comparing the information potential of diverse topological phases of matter.

*Satish Prajapati: ORCID iD: 0009-0006-3801-1137
†Satish Prajapati: iamsatish.gcect.ac@gmail.com

**ACKNOWLEDGMENTS**

The author received no financial support for this work.

*Satish Prajapati: ORCID iD: [0009-0006-3801-1137](0009-0006-3801-1137)

†Satish Prajapati: iamsatish.gcect.ac@gmail.com

# Supplemental Material

## From Qubits to Qumodes: Information Capacity of Anyonic Excitations

Satish Prajapati[*,†]
Department of Ceramic Technology, Government College of Engineering and Ceramic Technology
73, Abinash Chandra Banerjee Ln, Phool Bagan, Beleghata, Kolkata, West Bengal 700010

*Satish Prajapati: ORCID iD: [0009-0006-3801-1137](https://orcid.org/0009-0006-3801-1137)
†Contact author: iamsatish.gcect.ac@gmail.com

This document contains:
· Supplementary Note 1: Detailed Derivation of the Partition Function
· Supplementary Note 2: Maximum Entropy and the Uniform Distribution
· Supplementary Note 3: Finite-Temperature Analysis
· Supplementary Note 4: Connection to the Holevo Bound
· Supplementary Note 5: Extended Discussion on Experimental Realization
. Supplementary Note 6: Quantum Information Application: The Anyonic Qudit
· Supplementary Figures S1–S4
· Supplementary References

## I. Detailed Derivation of the Partition Function

This derivation assumes the grand canonical partition function for a single state factorizes, which holds for non-interacting particles or serves as an effective description for the statistical mechanics of Haldane exclusion statistics [19]. The partition function for a single quantum state with a maximum occupancy of m particles is defined by the sum over all allowed occupation numbers:

$$\mathcal{Z} = \sum_{n=0}^{m} e^{-\beta n(\epsilon - \mu)},$$

where $\beta = 1/(k_B T)$, is the energy of the state, and $\mu$ is the chemical potential. This is a finite geometric series. Using the identity for the sum of a geometric series,

$$\sum_{k=0}^{K} r^k = \frac{1 - r^{K+1}}{1 - r}, \quad \text{for } r \neq 1,$$

and setting $r = e^{-\beta(\epsilon - \mu)}$, we obtain:

$$\mathcal{Z} = \sum_{n=0}^{m} r^n = \frac{1 - r^{m+1}}{1 - r} = \frac{1 - e^{-\beta(m+1)(\epsilon - \mu)}}{1 - e^{-\beta(\epsilon - \mu)}}.$$

This is the general form used in the main text. The probability of occupancy n is given by the Boltzmann factor normalized by the partition function: $P(n) = e^{-\beta n(\epsilon - \mu)}/\mathcal{Z}$.

## II. Maximum Entropy and the Uniform Distribution

The von Neumann entropy $S = -\sum_{n=0}^{m} P(n) \log_2 P(n)$ is maximized when the probability distribution is uniform. We prove this using the method of Lagrange multipliers to maximize S under the constraint $\sum_{n=0}^{m} P(n) = 1$.

The Lagrangian is:

$$\Lambda = -\sum_n P(n) \ln P(n) + \lambda \left( \sum_n P(n) - 1 \right),$$

where we use natural logarithm for convenience (the base of the logarithm in the entropy definition only contributes a multiplicative constant, and the maximum is found at the same distribution). Taking the derivative with respect to P(n):

$$\frac{\partial \Lambda}{\partial P(n)} = -\ln P(n) - 1 + \lambda = 0.$$

This implies $\ln P(n) = \lambda - 1$ for all n, meaning all P(n) are equal. From the normalization constraint, with m+1 states, we find:

$$P(n) = \frac{1}{m+1} \quad \text{for all } n.$$

Substituting into the entropy formula yields the maximum capacity:

$$S_{\max} = -\sum_{n=0}^{m} \frac{1}{m+1} \log_2 \left(\frac{1}{m+1}\right) = \log_2(m+1).$$

Supplementary Figure S1 shows these uniform distributions for different statistics parameters g.

### III. Finite-Temperature Analysis

The main text focuses on the maximum capacity at $\epsilon = \mu$. Here, we analyze the entropy S as a function of $\beta(\epsilon - \mu)$ for different g values. The entropy is calculated from the full expression:

$$S = -\sum_{n=0}^{m} P(n) \log_2 P(n), \quad \text{where} \quad P(n) = \frac{e^{-\beta n(\epsilon - \mu)}}{Z}.$$

Supplementary Figure S2 shows S versus $\beta(\epsilon - \mu)$ for g = 1 (fermions), g = 1/2 (semions), and g = 1/3. The entropy peaks at $\epsilon = \mu (\beta(\epsilon - \mu) = 0))$, reaching its maximum value of $(\log_2(m+1)$. The width of the peak decreases as m increases, showing that systems with higher capacity are more sensitive to detuning from the chemical potential.

### IV. Connection to the Holevo Bound

The maximum entropy $\log_2(m+1)$ corresponds to the Holevo bound $\chi$ [16], which defines the ultimate classical information capacity of a quantum channel. For a quantum system that can be prepared in states $\rho_n$ with probabilities $p_n$, the bound is:

$$\chi = S\left(\sum_n p_n \rho_n\right) - \sum_n p_n S(\rho_n),$$

where $S(\rho)$ is the von Neumann entropy.

In our case, the "states" are the different occupation numbers n. For a quantum dot in the Coulomb blockade regime, these are energy eigenstates and are therefore orthogonal and perfectly distinguishable via a charge measurement. The Holevo bound thus simplifies to the Shannon entropy of the classical source:

$$\chi = -\sum_{n=0}^{m} p_n \log_2 p_n,$$

which is maximized by the uniform distribution, yielding $\chi_{\max} = \log_2(m+1)$. This confirms that our result is consistent with the fundamental limits of quantum information theory.

### V. Extended Discussion on Experimental Realization

The predicted conductance plateaus for the $\nu = 1/3$ state $((g = 1/3))$ can be observed using quantum dot spectroscopy [18]. Key experimental considerations:

Platform: A GaAs-based two-dimensional electron gas in the fractional quantum Hall regime.
Conditions: High magnetic field ($B \approx 10\, T$), low temperature (T < 100 mK) [9].

Expected Signals:

· Conductance dI/dV: Quantized plateaus as a function of gate voltage $V_g$. The number of plateaus (four) is the primary signature, corresponding to the discrete occupancies n = 0, 1, 2, 3. The plateau values are set by tunneling rates and are not expected to be precisely at integer multiples of $e^2/h$ [6]. The key prediction is the four-periodicity.

· Shot Noise: Peaks in noise power $S_I$ at transitions between plateaus, providing direct signatures of the fractional charge $e^*/e = 1/3$ tunneling [10,11]. while theoretical analyses of interferometers predict additional singular features in noise arising from anyonic tunneling processes [17].

This four-periodicity is consistent with recent experimental studies of anyonic Fabry-Pérot interferometers, which have observed oscillations with a period of 4 in the phase of the interference pattern, corresponding to the four possible occupation states of an anyon localized within the interferometer [17].

Supplementary Figure S3 shows simulated conductance and shot noise data, illustrating these expected signatures.

## VI. Quantum Information Application: The Anyonic Qudit

The quantization of entropy to ($S_{\max} = \log_2(m+1)$) bits, as derived in Supplementary Note 2, has a direct and profound implication: the anyonic state constitutes a native qudit—a higher-dimensional generalization of a qubit. For the (g = 1/3) state ($m = 3$), this corresponds to a four-level quantum system or "ququart," capable of encoding two bits of classical information.

### A. Qudit Readout Principle

The projective measurement of the qudit state is performed by a direct conductance measurement. The quantized conductance ($G \propto n$) serves as the pointer variable, projecting the system onto one of the four orthogonal charge occupancy eigenstates ($n = 0,1,2,3$). A single-shot measurement of ($G$) thus yields a direct readout of the two-bit state, a significant advantage over sequential measurements often required for multi-qubit systems.

### B. Hardware Implementation and Resource Analysis

Supplementary Figure S4 illustrates the proposed readout circuitry and contrasts it with the conventional approach.

- **Panel a (Anyonic Qudit):** The readout requires a single quantum dot tuned to the ($g = 1/3$) state, controlled by one plunger gate ($V_g$), with one pair of source (S) and drain (D) contacts, and a single analog-to-digital converter (ADC) for measurement.

- **Panel b (Two Qubits):** Encoding the same 4-dimensional Hilbert space with standard qubits requires two physically isolated quantum dots, two independent plunger gates ($V_{g1}, V_{g2}$), two pairs of contacts, and two separate measurement circuits (ADC1, ADC2).

### C. Distinguishability and the Orthogonality of States

A valid qudit requires its computational basis states to be distinguishable. In this system, the charge occupancy states ($|n\rangle$) are energy eigenstates (due to the large charging energy in the Coulomb blockade regime) and are therefore orthogonal ($(\langle n|n'\rangle = \delta_{nn'})$). A charge sensor (e.g., a quantum point contact or single-electron transistor) can distinguish between these states with high fidelity, fulfilling this requirement [20].

### D. Outlook towards Quantum Operations

While this work establishes the readout principle for a static anyonic qudit, performing quantum gate operations would require the controlled manipulation of superpositions of these charge states. This presents a fertile ground for future theoretical and experimental work, potentially leveraging microwave irradiation or non-adiabatic gate pulses for coherent control.

# Supplementary Figures

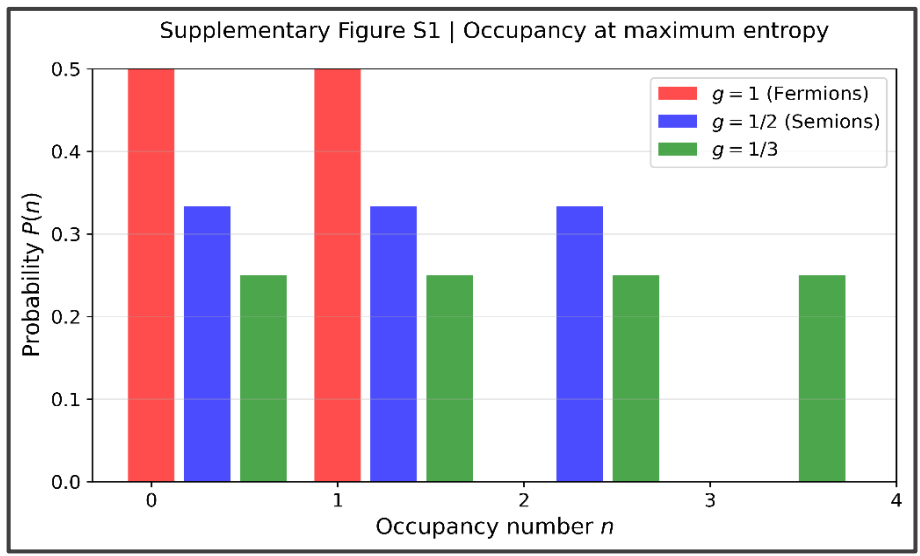

Supplementary Figure S1 | Occupancy probability distributions at maximum entropy. The probability distribution P(n) of finding n anyons in a single quantum state is shown for three different values of the exclusion statistics parameter g. At maximum entropy, which occurs when the energy level is aligned with the chemical potential ($\epsilon = \mu$)), all allowed occupational states are equally probable.

This results in a uniform distribution, and the corresponding maximum von Neumann entropy (information capacity) is ($S_{max} = \log_2(m+1)$, where $m = \lfloor 1/g \rfloor$ is the maximum allowed occupancy. For fermions ($g = 1, m = 1$), the capacity is 1 bit (qubit). For semions (($g = 1/2, m = 2$)) and the ($g = 1/3\ case\ ((m = 3))$, the capacities are approximately 1.585 bits and 2 bits, respectively.

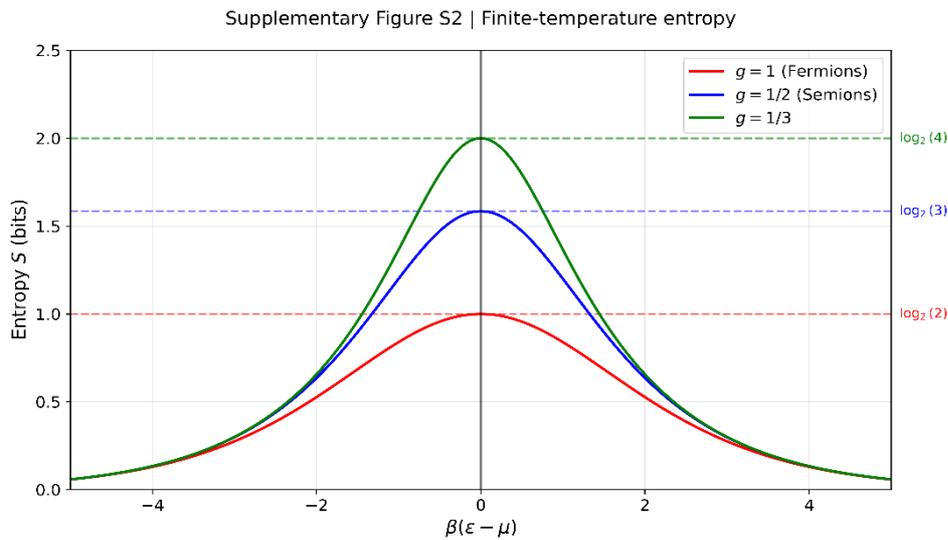

Supplementary Figure 2 | Finite-temperature dependence of the entropy. The von Neumann entropy S is plotted as a function of the detuning from the chemical potential, $\beta(\epsilon - \mu)$, for three different statistics parameters g. The entropy reaches its theoretical maximum, $S_{max} = \log_2(m+1)$ (indicated by dashed horizontal lines), only when the energy level is precisely tuned to the chemical potential ($\epsilon = \mu$)). The width of the entropy peak narrows as the maximum occupancy (m increases, indicating that systems with higher information capacity (e.g., ($g = 1/3$)) require more precise energy-level tuning to achieve their full capacity.

Supplementary Figure S3 | Simulated experimental signatures for $\nu = 1/3$ state

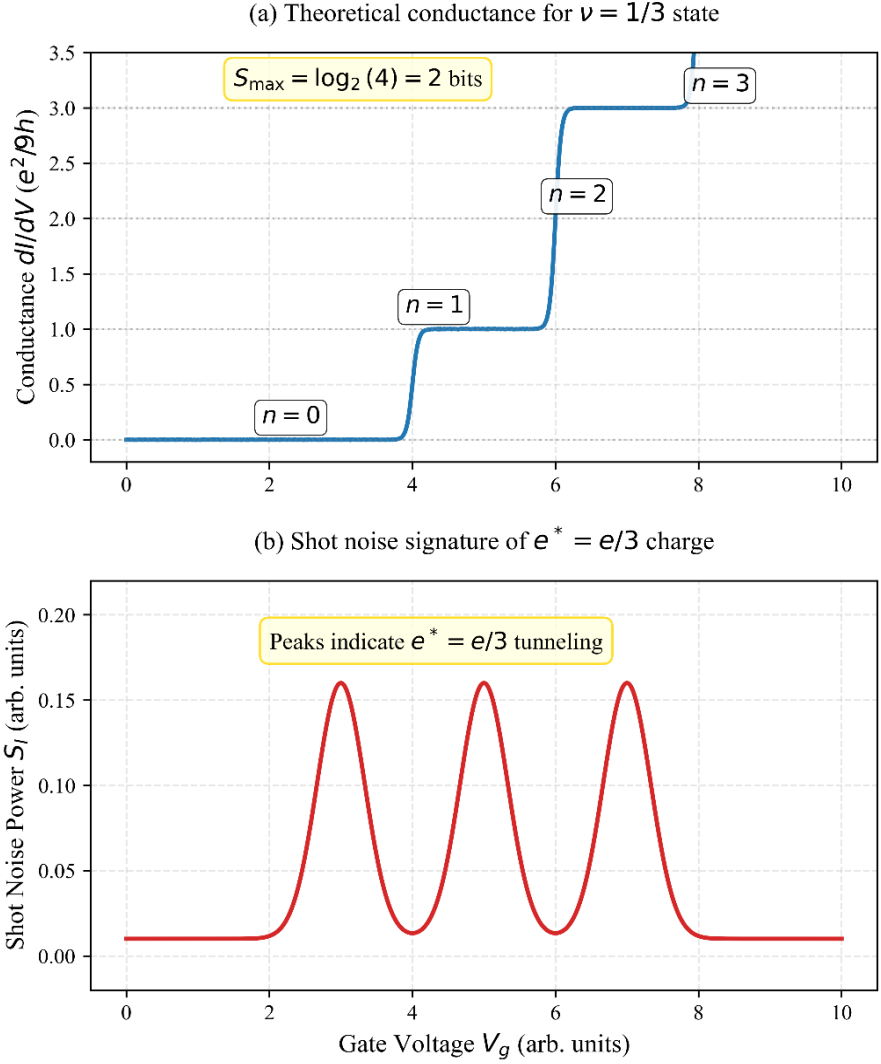

Supplementary Figure 3 | Simulated experimental signatures for the $\nu = 1/3$ state. (a) Theoretical prediction for the differential conductance $dI/dV$ through an anyon-trapping quantum dot as a function of gate voltage $V_g$. The four distinct plateaus correspond to the quantum dot being occupied by n = 0, 1, 2, 3 anyons (labeled), directly demonstrating the 2-bit information capacity predicted for statistics parameter g = 1/3. The conductance values are given in units of the fundamental quantum $e^2/9h$ for charge-$e/3$ quasiparticles. (b) The corresponding predicted shot noise power $S_I$. Peaks in the noise spectrum occur at the transitions between conductance plateaus and provide a signature of the fractional charge $e^* = e/3$ tunneling through the dot.

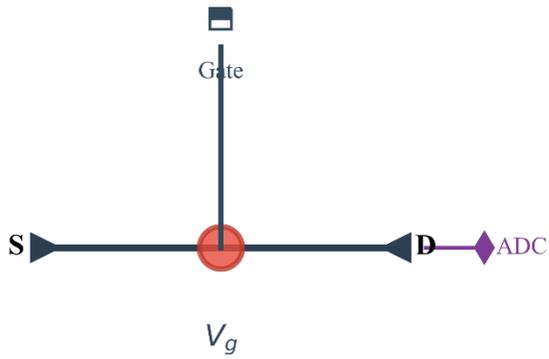
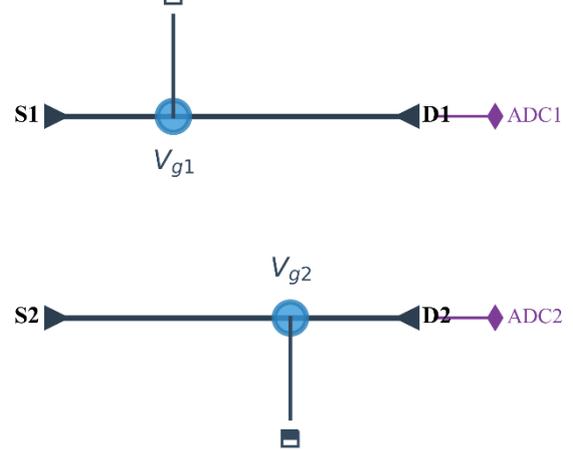

Supplementary Figure S4 | Quantum circuit implementation of an anyonic qudit. a, Measurement setup for a single anyonic qudit in the ( $g = 1/3$ ) state. A single quantized conductance measurement across the source (S) and drain (D) terminals, controlled by a plunger gate $(V_g)$, projects the state onto one of four charge occupancy states $((n = 0,1,2,3))$, encoding two bits of information. b, Equivalent setup for two conventional qubits required to span the same 4-dimensional Hilbert space, necessitating duplicate hardware.